\documentclass[twocolumn,prl]{revtex4-1}
\usepackage{amsmath}
\usepackage{amssymb}
\usepackage{graphicx}
\usepackage{esint}
\usepackage{bbm}
\usepackage{bm}
\usepackage{mathrsfs}

\usepackage{mathbbol}
\DeclareSymbolFontAlphabet{\amsmathbb}{AMSb}%

\newcommand{\bms}[1]{\mathbbmss{#1}}

\usepackage{hyperref}
\usepackage{color}

\definecolor{myblue}{RGB}{0,0,255}
\hypersetup{
    colorlinks,
    citecolor=myblue,
    linkcolor=myblue,
    urlcolor=myblue
}

\begin{document}
\title{Fluctuation Theorem Uncertainty Relation}
\author{Yoshihiko Hasegawa}
\email{hasegawa@biom.t.u-tokyo.ac.jp}
\affiliation{Department of Information and Communication Engineering, Graduate
School of Information Science and Technology, The University of Tokyo,
Tokyo 113-8656, Japan}
\author{Tan Van Vu}
\email{tan@biom.t.u-tokyo.ac.jp}
\affiliation{Department of Information and Communication Engineering, Graduate
School of Information Science and Technology, The University of Tokyo,
Tokyo 113-8656, Japan}
\date{\today}
\begin{abstract}
The fluctuation theorem is the fundamental equality in nonequilibrium
thermodynamics that is used to derive many important thermodynamic relations,
such as the second law of thermodynamics and the Jarzynski equality.
Recently, the thermodynamic uncertainty relation was discovered, which
states that the fluctuation of observables is lower bounded
by the entropy production. 
In the present Letter, we derive
a thermodynamic uncertainty relation from the
fluctuation theorem.
We refer to the obtained relation as the fluctuation theorem uncertainty relation,
and it is valid for arbitrary dynamics, stochastic as well as deterministic, and 
for arbitrary anti-symmetric observables for which a fluctuation theorem holds. 
We apply the fluctuation theorem uncertainty relation to an overdamped Langevin dynamics
for an anti-symmetric observable.
We demonstrate that the anti-symmetric 
observable satisfies the fluctuation theorem uncertainty relation,
but does not satisfy the relation reported for current-type observables
in continuous-time Markov chains. 
Moreover, we show that the fluctuation theorem uncertainty relation
can handle systems controlled by time-symmetric external protocols, in which the lower bound is
given by the work exerted on the systems.
\end{abstract}
\maketitle
\emph{Introduction.}---During the last two decades, stochastic thermodynamics
\cite{Ritort:NEArticle,Seifert:2012:FTReview,VandenBroeck:2015:Review}
accelerated the understanding of nonequilibrium systems through the
discovery of several thermodynamic relations. Among them, the fluctuation
theorem (\cite{Evans:1993:FT,Gallavotti:1995:Ensemble,Searles:1999:FT,Kurchan:1998:FT,Lebowitz:1999:LDP,Crooks:1999:CFT,Searles:2000:GeneralTFT,Jarzynski:2000:HamilFT,Gaspard:2004:FT}
and reviews \cite{Harris:2007:StocProcFT,Spinney:2013:FTReview})
is the central relation in nonequilibrium systems because this theorem leads to important
thermodynamic relations, such as the second law of thermodynamics,
the Green--Kubo relation \cite{Searles:2000:GKfromFT}, and the Jarzynski
equality \cite{Jarzynski:1997:Equality}, to name but a few. Recently,
a remarkable relation between fluctuation and the entropy production
was found, which is known as the thermodynamic uncertainty relation
(TUR) 
\cite{Barato:2015:UncRel,Barato:2015:FanoBound,Gingrich:2016:TUP,Polettini:2016:TUP,Pietzonka:2016:Bound,Hyeon:2017:TUR,Horowitz:2017:TUR,Proesmans:2017:TUR,Pietzonka:2017:FiniteTUR,Pigolotti:2017:EP,Garrahan:2017:TUR,Dechant:2018:TUR,Barato:2018:PeriodicTUR,Koyuk:2018:PeriodicTUR,Hwang:2018:TUR,Vroylandt:2019:PowerTradeOff,Dechant:2018:FRI,Terlizzi:2019:KUR,Hasegawa:2019:CRI,Dechant:2019:MTUR,Ito:2018:TimeTUR,Vu:2019:UTUR,Macieszczak:2018:TURLR}.
The TUR states that the fluctuation of observables, such
as the current, is lower bounded by the reciprocal of the entropy
production. The proof of the TUR has been carried out using the large
deviation principle \cite{Gingrich:2016:TUP,Polettini:2016:TUP,Pietzonka:2016:Bound,Horowitz:2017:TUR,Proesmans:2017:TUR,Pietzonka:2017:FiniteTUR,Barato:2018:PeriodicTUR,Koyuk:2018:PeriodicTUR,Vroylandt:2019:PowerTradeOff},
the fluctuation-response inequality \cite{Dechant:2018:FRI,Terlizzi:2019:KUR},
the Cram\'er--Rao inequality \cite{Hasegawa:2019:CRI,Dechant:2019:MTUR,Ito:2018:TimeTUR,Vu:2019:UTUR},
and the linear response around equilibrium \cite{Barato:2015:UncRel,Macieszczak:2018:TURLR}.
Although, as stated above, the fluctuation theorem can 
be used to derive many other thermodynamic
relations, the relations between the TUR and the fluctuation
theorem remain unclear. The universality of the fluctuation theorem
leads us to posit that the TUR can be derived through the fluctuation
theorem.

In the present Letter, we answer this question by obtaining the TUR for observables,
which are anti-symmetric under time reversal, from the fluctuation
theorem. We refer to the obtained relation as the fluctuation theorem uncertainty relation
(FTUR). Considering a detailed fluctuation theorem
with respect to the entropy production and the observable, we derive
the FTUR {[}see Eq.~\eqref{eq:main_result}{]}. As long as
the fluctuation theorem holds, the FTUR is valid for arbitrary systems
regardless of underlying dynamics and observables, and for arbitrary
observation time. 
Notably, the FTUR holds for deterministic dynamical ensembles, 
which cannot be handled by the abovementioned previous approaches. 
This is in contrast to existing TURs, which
assume particular stochastic dynamics (mostly Markovian), and their proofs were given for each dynamics.
The obtained results indicate that the TUR is a direct consequence of 
the fluctuation symmetry of the total entropy production. 
We apply the FTUR to the signum function of the current in an overdamped Langevin dynamics.
We show that the signum function of
the current does not satisfy the previously reported TUR {[}cf. Eq.~\eqref{eq:TUR_continuous}{]},
which holds for current-type observable in continuous-time Markov
chains.
Furthermore, the FTUR holds for systems controlled by time-symmetric external protocols.
In particular, when the systems are initially in equilibrium,
the FTUR holds with the total entropy production replaced by the work exerted on the systems. 
As an example of the FTUR with external protocols, we consider an overdamped dragged Brownian particle.

\emph{Model.}---We consider a system, which is continuous in space
and time, and assume that its time evolution is governed by a Markov
process. Although our description is based on continuous time and
continuous space, generalizations to discrete time or discrete space
are straightforward. 
We set the Boltzmann constant to unity. 
Let $x(t)$ be the position of the system at
time $t$ ($x(t)$ can be multidimensional), $\Gamma$ be a trajectory
from $t=0$ to $t=T$ ($T>0$), $\Gamma\equiv[x(t)]_{t=0}^{t=T}$, and $\Gamma^{\dagger}$
be its reversed trajectory, i.e., $\Gamma^{\dagger}\equiv[x(T-t)]_{t=0}^{t=T}$.
The system (i.e., the transition rate) can depend on an external protocol $\lambda(t)$. In the ensemble
level, the state of the system is depicted by $P(x,t)$, which is the
probability density that the system is in $x$ at time $t$. As is often
considered in stochastic thermodynamics, we consider forward and reverse
processes. We define $\mathcal{P}(\Gamma|x(0))$, the probability
of observing a trajectory $\Gamma$ in the forward process starting
from $x(0)$ at $t=0$, and $\mathcal{P}^{\dagger}(\Gamma^{\dagger}|x(T))$,
the probability of observing a trajectory $\Gamma^{\dagger}$ in the
reverse process starting from $x(T)$ at $t=T$. According to the
local detailed balance assumption, the total entropy production $\sigma(\Gamma)$
satisfies \cite{Seifert:2005:FT} $\sigma(\Gamma)=\ln\left[\mathcal{P}(\Gamma)/\mathcal{P}^{\dagger}(\Gamma^{\dagger})\right]$,
where $\mathcal{P}(\Gamma)\equiv P(x(0),0)\mathcal{P}(\Gamma|x(0))$
and $\mathcal{P}^{\dagger}(\Gamma^{\dagger})\equiv P(x(T),T)\mathcal{P}^{\dagger}(\Gamma^{\dagger}|x(T))$.
Throughout the present Letter, we consider cases in which $\sigma$ satisfies
the (strong) detailed fluctuation theorem $P(\sigma)/P(-\sigma)=e^{\sigma}$.
This condition is met when the system satisfies the following two
conditions: (i) the initial and final probability distributions agree,
$P(x,0)=P(x,T)$; and (ii) the external protocol is time symmetric, $\lambda(t)=\lambda(T-t)$
\cite{Spinney:2013:FTReview}. These conditions are typically satisfied
by systems in a steady state or in a periodic steady state with the periodic
protocol satisfying $\lambda(t)=\lambda(T-t)$. When (i) and (ii) are
satisfied, $\mathcal{P}(\Gamma)=\mathcal{P}^{\dagger}(\Gamma)$. Moreover,
satisfying (i) and (ii) implies that $\sigma(\Gamma)$ is anti-symmetric
under time reversal: 
\begin{equation}
\sigma(\Gamma^{\dagger})=-\sigma(\Gamma).\label{eq:sigma_anti_sym}
\end{equation}

Let $\phi(\Gamma)$ be an observable which is a function of $\Gamma$. Similar
to the total entropy production, we assume that $\phi(\Gamma)$ is
anti-symmetric under time reversal, i.e., 
\begin{equation}
\phi(\Gamma^{\dagger})=-\phi(\Gamma).\label{eq:anti_sym_def}
\end{equation}
As long as Eq.~\eqref{eq:anti_sym_def}
holds,
$\phi(\Gamma)$ can be an arbitrary function of $\Gamma$. The condition of Eq.~\eqref{eq:anti_sym_def} is typically
satisfied by the current, but there exist many other quantities that
can satisfy the condition.

Let $P(\sigma,\phi)$ be the probability that we observe the total entropy
production $\sigma$ and the observable $\phi$ in the forward process.
From Eqs.~\eqref{eq:sigma_anti_sym} and \eqref{eq:anti_sym_def},
we can show that $\sigma$ and $\phi$ obey the following strong detailed fluctuation
theorem \cite{GarciaGarcia:2010:UFT}:
\begin{align}
P(\sigma,\phi) & =\int\mathcal{D}\Gamma\,\delta(\sigma-\sigma(\Gamma))\delta(\phi-\phi(\Gamma))\mathcal{P}(\Gamma)\nonumber \\
 & =e^{\sigma}\int\mathcal{D}\Gamma^{\dagger}\,\delta(\sigma+\sigma(\Gamma^{\dagger}))\delta(\phi+\phi(\Gamma^{\dagger}))\mathcal{P}(\Gamma^{\dagger})\nonumber \\
 & =e^{\sigma}P(-\sigma,-\phi),\label{eq:FT_sigma_w_def}
\end{align}
where $\int \mathcal{D}\Gamma$ is the path integral. 

We now derive the FTUR solely from Eq.~\eqref{eq:FT_sigma_w_def}.
Reference~\citep{Merhav:2010:PropFromFT} examined the statistical properties
of entropy production from the fluctuation theorem. Inspired by Ref.~\citep{Merhav:2010:PropFromFT},
we introduce a probability density function $Q(\sigma,\phi)$ as follows: 
\begin{equation}
Q(\sigma,\phi)\equiv(1+e^{-\sigma})P(\sigma,\phi).\label{eq:Q_def}
\end{equation}
Here, $Q(\sigma,\phi)$ is normalized such that $\int_{0}^{\infty}d\sigma\int_{-\infty}^{\infty}d\phi\,Q(\sigma,\phi)=1$,
which directly follows from Eq.~\eqref{eq:FT_sigma_w_def} and 
$\int_{-\infty}^{\infty}d\sigma = \int_{-\infty}^{0}d\sigma + \int_{0}^{\infty}d\sigma$.
Then, $\left\langle \phi\right\rangle $ can be represented as the
expectation with respect to $Q(\sigma,\phi)$:
\begin{align}
    \left\langle \phi\right\rangle &\equiv\int_{-\infty}^{\infty}d\sigma\int_{-\infty}^{\infty}d\phi\,P(\sigma,\phi)\phi\nonumber\\
    &=\int_{0}^{\infty}d\sigma\int_{-\infty}^{\infty}d\phi\,P(\sigma,\phi)\phi(1-e^{-\sigma})\nonumber\\
    &=\left\langle \phi\tanh\left(\frac{\sigma}{2}\right)\right\rangle _{Q},\label{eq:w_as_f}
\end{align}
where $\left\langle \alpha(\sigma,\phi)\right\rangle _{Q}\equiv\int_{0}^{\infty}d\sigma\int_{-\infty}^{\infty}d\phi\,Q(\sigma,\phi)\alpha(\sigma,\phi)$
for arbitrary function $\alpha(\sigma,\phi)$.
Equation~\eqref{eq:w_as_f} holds for any observable $\phi(\Gamma)$ that is anti-symmetric under
time reversal [Eq.~\eqref{eq:anti_sym_def}]. 
Similarly, $\langle \sigma \rangle$ and $\left\langle \phi^{2}\right\rangle $ are
\begin{align}
    \left\langle \sigma\right\rangle &\equiv\int_{-\infty}^{\infty}d\sigma\int_{-\infty}^{\infty}d\phi\,P(\sigma,\phi)\sigma=\left\langle \sigma\tanh\left(\frac{\sigma}{2}\right)\right\rangle _{Q},\label{eq:sigma_as_g}\\
    \left\langle \phi^{2}\right\rangle &\equiv\int_{-\infty}^{\infty}d\sigma\int_{-\infty}^{\infty}d\phi\,P(\sigma,\phi)\phi^{2}=\left\langle \phi^{2}\right\rangle _{Q}.\label{eq:w2_def}
\end{align}
Applying the Cauchy--Schwarz inequality to Eq.~\eqref{eq:w_as_f},
we obtain 
\begin{align}
\left\langle \phi\right\rangle ^{2}=\left\langle \phi\tanh\left(\frac{\sigma}{2}\right)\right\rangle _{Q}^{2}\le\left\langle \phi^{2}\right\rangle _{Q}\left\langle \tanh\left(\frac{\sigma}{2}\right)^{2}\right\rangle _{Q}.\label{eq:w_expect}
\end{align}
Next, we want to show the following series of inequalities: 
\begin{equation}
\left\langle \tanh\left(\frac{\sigma}{2}\right)^{2}\right\rangle _{Q}\le\left\langle \tanh\left[\frac{\sigma}{2}\tanh\left(\frac{\sigma}{2}\right)\right]\right\rangle _{Q}\le\tanh\left(\frac{\left\langle \sigma\right\rangle }{2}\right).\label{eq:tanh_ineq}
\end{equation}
In order to show the first inequality part in Eq.~\eqref{eq:tanh_ineq}, we
define $\Delta(\sigma)\equiv\frac{\sigma}{2}\tanh\left(\frac{\sigma}{2}\right)-\mathrm{atanh}\left[\tanh\left(\frac{\sigma}{2}\right)^{2}\right]$.
We find that $\Delta(0)=0$ and $\Delta^{\prime}(\sigma)=(\sigma-\tanh(\sigma))/(2+2\cosh(\sigma))\ge0$
for $\sigma\ge0$, which shows $\Delta(\sigma)\ge0$ for $\sigma\ge0$
(note that the integration of $\langle\cdots\rangle_{Q}$ with respect
to $\sigma$ is in $[0,\infty)$, and thus we only have to consider the
$\sigma\ge0$ domain). Since $\mathrm{tanh}(\sigma)$ is a strictly
increasing function, we prove the first inequality in Eq.~\eqref{eq:tanh_ineq}.
The second inequality part in Eq.~\eqref{eq:tanh_ineq} can be
proved as follows. Since $\tanh(\sigma)$ is a concave function
for $\sigma\ge0$, by using the Jensen inequality, we find 
$\left\langle \tanh\left[\frac{\sigma}{2}\tanh\left(\frac{\sigma}{2}\right)\right]\right\rangle _{Q}\le\tanh\left[\frac{1}{2}\left\langle \sigma\tanh\left(\frac{\sigma}{2}\right)\right\rangle _{Q}\right]$, which proves the second inequality part in Eq.~\eqref{eq:tanh_ineq}
by using Eq.~\eqref{eq:sigma_as_g}. Combining Eqs.~\eqref{eq:w2_def} through \eqref{eq:tanh_ineq}, we obtain 
$\langle \phi^2 \rangle / \langle \phi \rangle^2 \ge \tanh(\langle \sigma \rangle / 2)^{-1}$,
which yields
\begin{equation}
\frac{\mathrm{Var}[\phi]}{\left\langle \phi\right\rangle ^{2}}\ge\frac{2}{e^{\left\langle \sigma\right\rangle }-1}.\label{eq:main_result}
\end{equation}
Here, $\mathrm{Var}[\phi]\equiv\langle\phi^{2}\rangle-\langle\phi\rangle^{2}$
is the variance of $\phi$. We refer to Eq.~\eqref{eq:main_result}
as the FTUR, which is the main result of the present Letter.

We make some remarks on Eq.~\eqref{eq:main_result}. 
Equation~\eqref{eq:main_result} is valid for arbitrary dynamics as long as the 
fluctuation theorem of Eq.~\eqref{eq:FT_sigma_w_def} holds.
Therefore, Eq.~\eqref{eq:main_result} can be applied to continuous-, as well as discrete-time, Markov chains. 
Indeed, the expression
of Eq.~\eqref{eq:main_result} is equivalent to the bound obtained
for discrete-time Markov chains \citep{Proesmans:2017:TUR}. 
The bound of Eq.~\eqref{eq:main_result} is always smaller than that
of the well-known TUR 
\begin{equation}
\frac{\mathrm{Var}[\phi]}{\left\langle \phi\right\rangle ^{2}}\ge\frac{2}{\left\langle \sigma\right\rangle },\label{eq:TUR_continuous}
\end{equation}
which is valid for continuous-time Markov chains. 
Discrete-time Markov chains do not satisfy Eq.~\eqref{eq:TUR_continuous} 
\cite{Shiraishi:2017:TURViolate,Proesmans:2017:TUR}.
The bound of Eq.~\eqref{eq:TUR_continuous}
has been proved for current-type observables and for the first-passage
time (the former case was proved for a finite-time case). Still,
as will be demonstrated, Eq.~\eqref{eq:TUR_continuous}
is not satisfied even in continuous-time Markov chains when we consider
an observable other than the current.

The quantity $\phi(\Gamma)$ can be arbitrary as long as Eq.~\eqref{eq:anti_sym_def}
holds. This condition is typically satisfied by the current, but can be satisfied
by other quantities as well. Let $\jmath(\Gamma)$ be the current,
which can of course satisfy $\jmath(\Gamma^{\dagger})=-\jmath(\Gamma)$.
Then any observable $h(\jmath(\Gamma))$, where $h(x)$ is an arbitrary
odd function, satisfies $h(\jmath(\Gamma^{\dagger}))=-h(\jmath(\Gamma))$,
and thus the FTUR holds for $h(\jmath(\Gamma))$ (this case is
considered in the example section). Moreover, $\sigma$ can be an
observable other than the total entropy production. Although, for clarity, we have
assumed that $\sigma$ is the total entropy production in Eq.~\eqref{eq:FT_sigma_w_def},
any set of observables $\sigma$ and $\phi$ that satisfy
the fluctuation theorem of Eq.~\eqref{eq:FT_sigma_w_def}
admit the FTUR of Eq.~\eqref{eq:main_result}.
In particular, we can obtain the FTUR for which the thermodynamic cost is the work exerted on the system.
Suppose that the initial distributions for both the forward and reverse
processes are equilibrium distributions. 
Then, $w$, the work exerted
on the system, satisfies the Crooks work relation \cite{Crooks:1999:CFT,Crooks:2000:PathEnsemble}
$P(w)/P^{\dagger}(-w)=e^{(w-\Delta F)/\mathcal{T}}$, where $\mathcal{T}$
is temperature, $P^{\dagger}(-w)$ is the probability of observing $-w$
in the reverse process, and $\Delta F$ is the free energy difference
between equilibrium distributions corresponding to $\lambda(T)$ and
$\lambda(0)$. Furthermore, when a symmetric external protocol $\lambda(t)=\lambda(T-t)$
is applied, the forward and reverse processes are indistinguishable,
and the free energy difference vanishes, $\Delta F=0$, resulting in $P(w)/P(-w)=e^{w/\mathcal{T}}$. Therefore,
under these conditions,
any observables $\phi(\Gamma)$ satisfying
Eq.~\eqref{eq:anti_sym_def} obey the following FTUR: 
\begin{equation}
\frac{\mathrm{Var}[\phi]}{\left\langle \phi\right\rangle ^{2}}\ge\frac{2}{e^{\left\langle w\right\rangle /\mathcal{T}}-1}.\label{eq:GTUR_work}
\end{equation}

Thus far, we have been concerned with stochastic systems. 
Historically, the fluctuation theorem was first demonstrated on deterministic dynamical ensembles
\cite{Evans:1993:FT}.
We can show that the FTUR also holds in such deterministic systems. 
Consider an $N$-particle system, where $\bm{q}_{i}(t)$ and $\bm{p}_{i}(t)$
denote the coordinates and the momenta of the $i$th particle at time $t$.
Let $\mathbb{\Gamma}(t)\equiv [\bm{q}(t),\bm{p}(t)] \equiv[\bm{q}_{1}(t),...,\bm{q}_{N}(t),\bm{p}_{1}(t),...,\bm{p}_{N}(t)]$
be a point in a phase space at time $t$,
and
let $\rho(\mathbb{\Gamma},t)$
be the distribution function of the phase space at time $t$.
The time evolution of $\mathbb{\Gamma}(t)$
is governed by the deterministic differential equation of $\dot{\mathbb{\Gamma}}$ 
(the overdot denotes the time derivative),
which is assumed to be time reversible so that the conjugate dynamics exists. 
We assume
that the initial ensemble ($t=0$) obeys a given distribution
(e.g., equilibrium distribution) and, for $t>0$, a constant field is applied
to the system. 
We define a dissipation from $t=0$ to $t=T$  as
$\Sigma \equiv\ln\left({\rho(\mathbb{\Gamma}(0),0)}/{\rho(\mathbb{\Gamma}(T),0)}\right)-\int_{0}^{T}\Upsilon(\mathbb{\Gamma}(s))ds$ \cite{Searles:2000:GeneralTFT}, where $\Upsilon(\mathbb{\Gamma})\equiv ({\partial}/{\partial \mathbb{\Gamma}})\cdot \dot{\mathbb{\Gamma}}$
is the phase space compression factor \cite{Evans:2008:NoneqLiquid} (the dot ``$\cdot$'' denotes the inner product). 
It is known that $\Sigma$ satisfies the fluctuation theorem, $P(\Sigma)/P(-\Sigma)=e^\Sigma$,
under mild conditions on the initial ensemble and dynamics
\cite{Searles:2000:GeneralTFT, Evans:2008:NoneqLiquid}. 
Analogous to Eq.~\eqref{eq:anti_sym_def},
we consider an arbitrary observable $\Phi$ defined from $t=0$ to $t=T$, which is assumed to be
anti-symmetric under time reversal. 
Extending the derivation of Ref.~\cite{Searles:2000:GeneralTFT}, 
we can show that the fluctuation theorem 
$P(\Sigma,\Phi)/P(-\Sigma,-\Phi)=e^\Sigma$
holds in the deterministic dynamical ensembles,
which indicates the satisfaction of the FTUR 
 (see \cite{Supp:PhysRev} for details of the derivation and numerical verification). 

\nocite{Morriss:1998:Thermostats}
\nocite{Tuckwell:2009:HHeq}

We next discuss the equality condition of Eq.~\eqref{eq:main_result}.
When the equality is attained in both Eqs.~\eqref{eq:w_expect} and
\eqref{eq:tanh_ineq}, the equality of the FTUR is satisfied. According
to the equality condition of the Cauchy--Schwarz inequality, the equality
of Eq.~\eqref{eq:w_expect} is satisfied only when $\phi\propto\tanh(\sigma/2)$.
The first inequality part in Eq.~\eqref{eq:tanh_ineq} becomes equality
only at $\sigma=0$. From the equality condition of the Jensen inequality,
the second inequality part in Eq.~\eqref{eq:tanh_ineq} saturates
only when $\tanh(\sigma)$ is a linear function, which is asymptotically
achieved for $\sigma\to0$. Combining all of these conditions, we find
that the equality of Eq.~\eqref{eq:main_result} is asymptotically
satisfied if and only if $\phi\propto\sigma$ and $\sigma\to0$. When
$\sigma\to0$, the system reduces to equilibrium. It has been reported
that the total entropy production satisfies the equality of the TUR
near equilibrium \cite{Gingrich:2016:TUP,Pigolotti:2017:EP,Macieszczak:2018:TURLR,Hasegawa:2019:CRI}, which agrees with our equality condition.

\begin{figure}
\includegraphics[width=8cm]{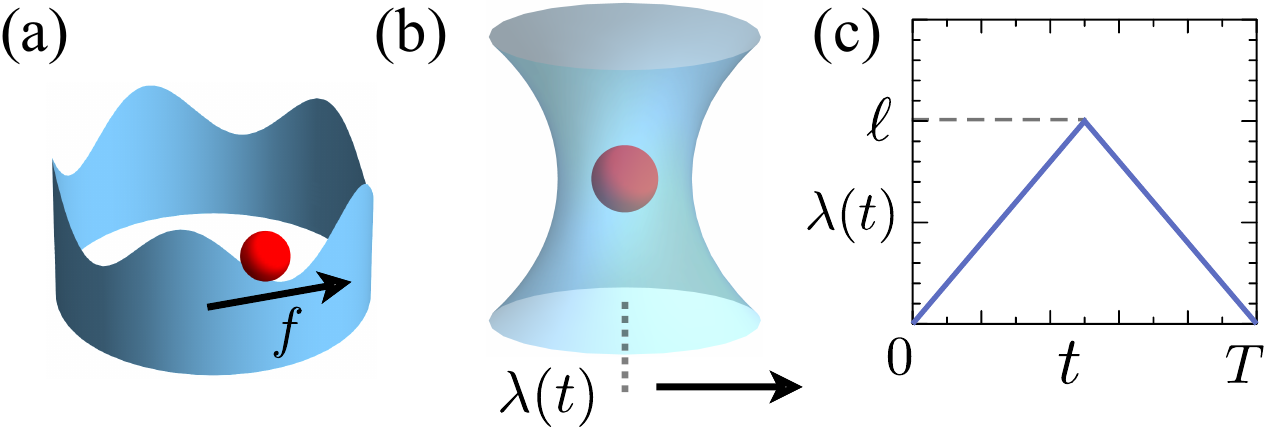}

\caption{Models considered in examples 1 and 2. (a) Particle on a ring topology with the drift term $A(x)$ in example 1. (b) Dragged Brownian particle, where the potential is 
manipulated by a protocol $\lambda(t)$ in example 2. (c)
Protocol $\lambda(t)$, defined by Eq.~\eqref{eq:lambda_def}, as a function of $t$, which is applied to the dragged Brownian particle shown by (b). 
$\ell$ denotes the height of the protocol. 
\label{fig:model}}
\end{figure}

\emph{Example 1.}---We apply the FTUR to an overdamped particle on
a ring (Fig.~\ref{fig:model}(a)), which has been extensively investigated in the
literature \cite{Speck:2006:RingModel,VandenBroeck:2010:ThreeFace2}. 
Without loss of generality, we assume that the circumference of the ring is $1$. 
We consider 
\begin{equation}
\dot{x}=A(x)+\sqrt{2D}\xi(t),\label{eq:Langevin_def}
\end{equation}
where $A(x)$ is a periodic drift function ($A(x)=A(x+1)$), $D>0$
is the noise intensity, and $\xi(t)$ is the white Gaussian noise
with $\langle\xi(t)\rangle=0$ and $\langle\xi(t)\xi(t')\rangle=\delta(t-t')$.
Let $P(x,t)$ be the probability density of $x$ at time $t$. 
The Fokker--Planck equation of Eq.~\eqref{eq:Langevin_def} is $\partial_{t}P(x,t)=-\partial_{x}J(x,t)$,
where $J(x,t)\equiv A(x)P(x,t)-D\partial_{x}P(x,t)$ is the probability
current. 
We use the following generalized current: $\jmath(\Gamma)\equiv\int_{0}^{T}\Lambda(x)\circ\dot{x}dt$,
 where $\circ$ is the Stratonovich product, and $\Lambda(x)$ is a
projection function. 
We consider observables defined by
$\phi_{\mathrm{sgn}}(\Gamma)\equiv\mathrm{sign}(\jmath(\Gamma))$, where
$\mathrm{sign}(x)$ is the signum function. $\phi_{\mathrm{sgn}}$ simply returns the sign of $\jmath$.
Since $\mathrm{sign}(x)$ is an odd function, $\phi_{\mathrm{sgn}}$
obeys the FTUR of Eq.~\eqref{eq:main_result}.

We explicitly calculate $\mathrm{Var}[\phi_\mathrm{sgn}]$ and $\langle\phi_\mathrm{sgn}\rangle$.
We consider $A(x)=f$, where $f$ is a constant force applied to the particle. 
Since we consider a ring of circumference of $1$, $P(x,t) \to 1$ for $t \to \infty$. 
Therefore, the steady-state current is $J^\mathrm{ss} = f$.
We use $\Lambda(x)=1$ in $\jmath(\Gamma)$, with which the current simply
gives the position at time $t=T$ 
on the infinite line, $\jmath(\Gamma)=x(T) - x(0)$. Furthermore, on the infinite line,
$P(x,t)$
is a Gaussian distribution with the mean $ft$ and the variance
$2Dt$ when $x(0)=0$ \cite{Risken:1989:FPEBook}. Let $\bms{P}(\phi_{\mathrm{sgn}},t)$
be the probability of $\phi_{\mathrm{sgn}}\in\{-1,1\}$ at time $t$, which
is expressed by 
$\bms{P}(1,t)=\frac{1}{2}\left[1+\mathrm{erf}\left(\frac{f}{2}\sqrt{\frac{t}{D}}\right)\right]$ and $\bms{P}(-1,t)=\frac{1}{2}\left[1-\mathrm{erf}\left(\frac{f}{2}\sqrt{\frac{t}{D}}\right)\right]$.
The explicit expression of $\bms{P}(\phi_\mathrm{sgn},t)$
confers $\mathrm{Var}[\phi_{\mathrm{sgn}}]/\left\langle \phi_{\mathrm{sgn}}\right\rangle ^{2}=-1+\mathrm{erf}\left(f\sqrt{t/D}/2\right)^{-2}$.
Since the entropy production from $t=0$ to $t=T$ is given by $\left\langle \sigma\right\rangle =T\int_0^1 dx\,D^{-1}A(x)J^{\mathrm{ss}}=Tf^{2}/D$,
we obtain \cite{Supp:PhysRev}
\begin{equation}
\frac{\mathrm{Var}[\phi_{\mathrm{sgn}}]}{\left\langle \phi_{\mathrm{sgn}}\right\rangle ^{2}}=-1+\mathrm{erf}\left(\frac{\sqrt{\langle\sigma\rangle}}{2}\right)^{-2}.\label{eq:ratio_sign}
\end{equation}
The right-hand side of Eq.~\eqref{eq:ratio_sign} is larger than
the lower bound of Eq.~\eqref{eq:main_result}, $-1+\mathrm{erf}\left(\sqrt{\langle\sigma\rangle}/2\right)^{-2}\ge2/[e^{\left\langle \sigma\right\rangle }-1]$.
This relation is obvious when evaluating both sides numerically,
but we provide a proof in \cite{Supp:PhysRev}. 

\begin{figure}
\includegraphics[width=8.5cm]{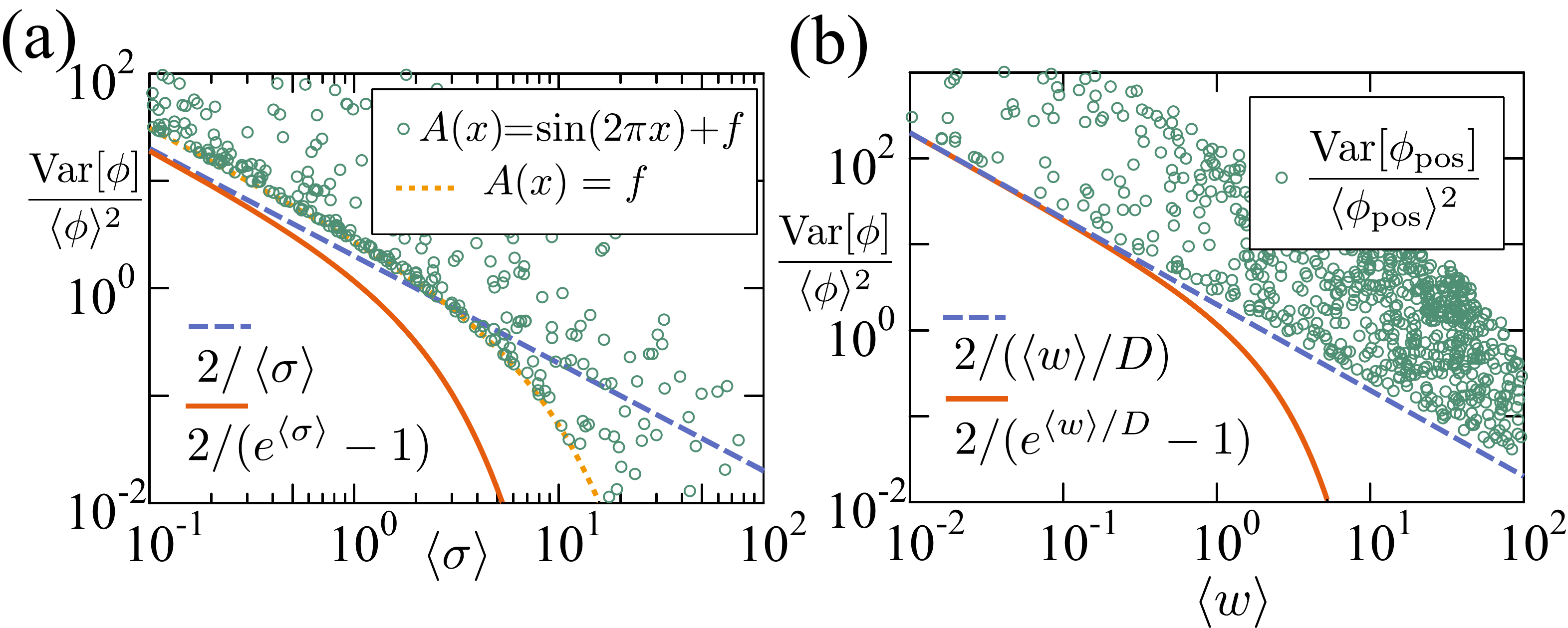}
\caption{$\mathrm{Var}[\phi]/\langle\phi\rangle^{2}$ and the lower bounds
of the TUR and FTUR
as a function $\langle\sigma\rangle$
(or $\langle w \rangle$). The lower bounds of Eqs.~\eqref{eq:main_result} and \eqref{eq:TUR_continuous} are depicted by solid and dashed lines, respectively.
(a) 
Results of the particle on a ring in example 1. 
The dotted line and circles denote
$\mathrm{Var}\left[\phi_{\mathrm{sgn}}\right]/\left\langle \phi_{\mathrm{sgn}}\right\rangle ^{2}$ for $A(x)=f$ and $A(x)=\sin(2\pi x)+f$, respectively, 
where $f$, $T$, and $D$ are randomly selected from $f\in[0.01,3.0]$, $T\in[0.1,3.0]$, and $D\in[0.01,1.0]$.
(b) Results of the dragged Brownian particle in example 2. 
Circles denote $\mathrm{Var}[\phi_\mathrm{pos}]/\langle\phi_\mathrm{pos} \rangle^{2}$
for randomly selected $\beta$, $\ell$, and $T$ with $D=1$. $\beta$, $\ell$, and $T$
are selected as $\beta \in [0.01,10.0]$, $\ell \in [0.01,10.0]$, and $T \in [0.01,10.0]$. 
\label{fig:experiment}}
\end{figure}

We plot $\mathrm{Var}\left[\phi_{\mathrm{sgn}}\right]/\left\langle \phi_{\mathrm{sgn}}\right\rangle ^{2}$
{[}Eq.~\eqref{eq:ratio_sign}{]} in Fig.~\ref{fig:experiment}(a) for $A(x)=f$ and $A(x)=\sin(2\pi x) + f$.
For $A(x)=f$, Eq.~\eqref{eq:ratio_sign} is depicted by a dotted line, and
the lower bounds of Eq.~\eqref{eq:main_result}
and Eq.~\eqref{eq:TUR_continuous} are shown by solid and dashed
lines, respectively. 
Although Eq.~\eqref{eq:ratio_sign}
is larger than the bound of Eq.~\eqref{eq:main_result}, it 
does not satisfy Eq.~\eqref{eq:TUR_continuous}, which indicates that the
continuous TUR {[}Eq.~\eqref{eq:TUR_continuous}{]} does not generally
hold for quantities that are anti-symmetric under time reversal.
We check the inequality for $A(x)=\sin(2\pi x) + f$,
which has a non-Gaussian distribution, with computer simulation. 
We randomly select $f$, $T$, and $D$, and calculate $\mathrm{Var}\left[\phi_{\mathrm{sgn}}\right]/\left\langle \phi_{\mathrm{sgn}}\right\rangle ^{2}$ and $\langle \sigma\rangle$ for the selected parameter values
as the average of $10^6$ trajectories
(the range of the parameters is shown in the caption of Fig.~\ref{fig:experiment}(a)),
and the realizations are shown by circles in Fig.~\ref{fig:experiment}(a).
We can see that $\mathrm{Var}\left[\phi_{\mathrm{sgn}}\right]/\left\langle \phi_{\mathrm{sgn}}\right\rangle ^{2}$ for $A(x)=\sin(2\pi x) + f$ is larger than the result of $A(x)=f$,
indicating that the case of $A(x)=f$ appears to be the lower bound case of this particular example. 
We again confirm that the conventional TUR is not satisfied for larger $\langle \sigma \rangle$. 

\emph{Example 2.}---Next, we consider an overdamped dragged Brownian particle (Fig.~\ref{fig:model}(b)) \cite{Wang:2002:DraggedBrown,Zon:2003:DraggedBrownian}
to test the FTUR of Eq.~\eqref{eq:GTUR_work}.
The dragged Brownian particle is important in
stochastic thermodynamics from both theoretical and experimental viewpoints. 
We consider the following
Langevin equation: $\dot{x}=-\partial_x U(x,\lambda(t))+\sqrt{2D}\xi(t)$,
where $U(x,\lambda) \equiv \beta(x-\lambda)^{2}/2$ is a potential function
($\beta>0$ is a model parameter), $\lambda(t)$ is an external
protocol, and $\xi(t)$ and $D$ are the same as in Eq.~\eqref{eq:Langevin_def}. We consider a time-symmetric protocol defined by 
\begin{equation}
\lambda(t)=\begin{cases}
\frac{2\ell}{T}t & 0\le t<\frac{T}{2}\\
-\frac{2\ell}{T}t+2\ell & \frac{T}{2}\le t\le T
\end{cases},\label{eq:lambda_def}
\end{equation}
where $\ell$ is the height of the signal (Fig.~\ref{fig:model}(c)). $\lambda(t)$
of Eq.~\eqref{eq:lambda_def} satisfies time symmetry, $\lambda(T-t)=\lambda(t)$.
Suppose that the system is in equilibrium at $t=0$, $P(x,0)=P^{\mathrm{eq}}(x,\lambda(0))$,
where $P^{\mathrm{eq}}(x,\lambda)\equiv \mathscr{N}\exp\left(-U(x,\lambda)/D\right)$
is the equilibrium distribution corresponding to $\lambda$ ($\mathscr{N}$ is a normalization constant).
We consider an observable
$\phi_{\mathrm{pos}}(\Gamma)\equiv\jmath(\Gamma)$ with $\Lambda(x) = 1$.
$\phi_{\mathrm{pos}}(\Gamma)$ simply gives the position of the particle at time $t=T$,
$\phi_{\mathrm{pos}}(\Gamma)=x(T)-x(0)$. Based on these assumptions,
$\phi_{\mathrm{pos}}(\Gamma)$ satisfies the FTUR given by Eq.~\eqref{eq:GTUR_work}
with $\mathcal{T}$ replaced by $D$.
In the dragged Brownian particle, the work $w$ exerted on the particle is
given by \cite{Seifert:2012:FTReview} $w(\Gamma)=\int_{0}^{T}dt\,\partial_{\lambda}U(x,\lambda)\dot{\lambda}$.
Since the probability density $P(x,t)$
is a Gaussian distribution for all $t$ and $\lambda(t)$ is a piecewise linear function [Eq.~\eqref{eq:lambda_def}], $\mathrm{Var}[\phi_{\mathrm{pos}}]/\langle\phi_{\mathrm{pos}}\rangle^{2}$
and $\langle w \rangle$ can be calculated analytically
\cite{Supp:PhysRev}:
\begin{align}
    \frac{\mathrm{Var}[\phi_{\mathrm{pos}}]}{\left\langle \phi_{\mathrm{pos}}\right\rangle ^{2}}&=\frac{D\beta T^{2}\left(1-e^{-\beta T}\right)}{2\ell^{2}\left(e^{-\beta T}-2e^{-\beta T/2}+1\right)^{2}},\label{eq:var_mean_dragged}\\
    \left\langle w\right\rangle &=\frac{4\ell^{2}\left(-e^{-\beta T}+\beta T+4e^{-\beta T/2}-3\right)}{\beta T^{2}}.\label{eq:w_average}
\end{align}
We randomly select $\beta$, $T$, and $\ell$ with $D=1$, and
calculate $\mathrm{Var}(\phi_{\mathrm{pos}})/\langle\phi_{\mathrm{pos}}\rangle^{2}$
and the average work $\langle w \rangle$ for the selected parameter values (the
range of the parameters is shown in the caption of Fig.~\ref{fig:experiment}(b)).
We repeat this calculation many times and plot $\mathrm{Var}(\phi_{\mathrm{pos}})/\langle\phi_{\mathrm{pos}}\rangle^{2}$
as a function of $\langle w \rangle$ in Fig.~\ref{fig:experiment}(b). 
$2/[e^{\langle w \rangle / D} - 1]$ and
$2/(\langle w\rangle/D)$ 
are depicted by solid and dashed
lines, respectively. 
We can confirm that all the realizations (circles) are above the bound of Eq.~\eqref{eq:GTUR_work},
indicating that Eq.~\eqref{eq:GTUR_work} holds for the dragged Brownian particle. 
Still, we can see that all the realizations are even above 
$2 / (\langle w\rangle / D)$ (dashed line).
This tighter bound is an analogue of Eq.~\eqref{eq:TUR_continuous} for the system
subject to the external protocol. 
Indeed, we can prove that $\mathrm{Var}[\phi_{\mathrm{pos}}]/\left\langle \phi_{\mathrm{pos}}\right\rangle ^{2}\ge2/(\left\langle w\right\rangle /D)$ and 
this inequality saturates when $\beta T\to 0$ \cite{Supp:PhysRev}.
This result induces us to conjecture that the FTUR of Eq.~\eqref{eq:GTUR_work}
has this tighter bound for general continuous-time systems
with equilibrium initial distributions and time-symmetric external protocols. 

We also tested the FTUR for a discrete-time random walk on a ring with
an observable counting the number of laps
and confirmed that the FTUR is satisfied for this system (see \cite{Supp:PhysRev}). 

\emph{Conclusion.}---In the present Letter, we have derived the FTUR solely from
the fluctuation theorem with respect to the total entropy production
and the observable, which is anti-symmetric under time reversal. 
Although the bound of the FTUR is weaker than that of the conventional TUR, 
the FTUR is general in the sense that it can handle systems
that have not been covered by the previously reported TURs. 
Since the fluctuation
theorem is the central relation in nonequilibrium thermodynamics,
the present study can be a basis for obtaining other thermodynamic bounds.

\emph{Acknowledgments.}---
We would like to thank Amos Maritan for providing valuable comments. 
The present study was supported by MEXT KAKENHI
Grant No.~JP16K00325, JP19K12153.

\end{document}


\title{Supplemental Material for \\``Fluctuation Theorem Uncertainty Relation''}
\author{Yoshihiko Hasegawa and Tan Van Vu}

\maketitle
This supplementary material describes the calculations introduced
in the main text. Equation and figure numbers are prefixed with S
(e.g., Eq.~(S1) or Fig.~S1). Numbers without this prefix (e.g.,
Eq.~(1) or Fig.~1) refer to items in the main text.

\section{Fluctuation theorem for deterministic dynamical ensembles}

\subsection{Derivation}

Our derivation of the TUR does not depend on the underlying dynamics
as long as the fluctuation theorem of Eq.~\FTUsigmaUwUdef{} holds. In the main
text, we considered stochastic dynamics for concreteness. Here, we also derive
the fluctuation theorem of Eq.~\FTUsigmaUwUdef{} for deterministic dynamical ensembles.

We first review the fluctuation theorem with respect to the dissipation
alone, following the derivation of Ref.~\cite{Searles:2000:GeneralTFT}.
Reference~\cite{Searles:2000:GeneralTFT} provided a proof for general
deterministic dynamical ensembles.
Consider an $N$-particle system, where $\bm{q}_{i}(t)$ and $\bm{p}_{i}(t)$
denote the coordinates and the momenta of the $i$th particle at time $t$. We assume
that the initial ensemble ($t=0$) obeys a given distribution
(e.g., equilibrium distribution) and, for $t>0$, a constant field is applied
to the system. Let $\mathbb{\Gamma}(t)\equiv [\bm{q}(t),\bm{p}(t)] \equiv[\bm{q}_{1}(t),...,\bm{q}_{N}(t),\bm{p}_{1}(t),...,\bm{p}_{N}(t)]$
be a point in a phase space at time $t$. The time evolution of $\mathbb{\Gamma}(t)$
is governed by a deterministic differential equation with respect to $\dot{\mathbb{\Gamma}}$,
which is assumed to be reversible so that the conjugate dynamics exists.
We are interested in observables at time $t=T$ ($T>0$). Let $\rho(\mathbb{\Gamma},t)$
be the distribution function of the phase space at time $t$. It is known that
$\rho(\mathbb{\Gamma},t)$ obeys the Liouville equation:
\begin{equation}
\frac{d}{dt}\rho(\mathbb{\Gamma}(t),t)=-\rho(\mathbb{\Gamma}(t),t)\Upsilon(\mathbb{\Gamma}(t)),\label{eq:dfdt_def}
\end{equation}
where ${\displaystyle \Upsilon(\mathbb{\Gamma})\equiv\frac{\partial}{\partial\mathbb{\Gamma}}\cdot\dot{\mathbb{\Gamma}}}$
is the phase space compression factor ($\cdot$ denotes the inner product).
For Hamiltonian dynamics, the phase space compression factor vanishes, i.e.,
$\Upsilon(\mathbb{\Gamma})=0$. From Eq.~(\ref{eq:dfdt_def}), we
obtain
\begin{equation}
\rho(\mathbb{\Gamma}(T),T)=\exp\left[-\int_{0}^{T}\Upsilon(\mathbb{\Gamma}(s))ds\right]\rho(\mathbb{\Gamma}(0),0),\label{eq:rho_formal_sol}
\end{equation}
which is known as the Lagrangian form of the Kawasaki distribution
function \cite{Evans:2008:NoneqLiquid}. Let $\Sigma$ be a
dissipation defined by
\begin{equation}
\Sigma \equiv\ln\left(\frac{\rho(\mathbb{\Gamma}(0),0)}{\rho(\mathbb{\Gamma}(T),0)}\right)-\int_{0}^{T}\Upsilon(\mathbb{\Gamma}(s))ds.\label{eq:Sigma_def}
\end{equation}
For instance, in isokinetic (isothermal) thermostated systems, $\Sigma$ corresponds to the dissipative flux,
which  is the adiabatic derivative of the internal energy \cite{Morriss:1998:Thermostats,Searles:2000:GeneralTFT,Evans:2008:NoneqLiquid}.
Since the system
is deterministic, $\Sigma$ can be uniquely specified given either
the initial point $\mathbb{\Gamma}(0)$ at time $t=0$ or the end
point $\mathbb{\Gamma}(T)$ at time $t=T$.
Given a point $\mathbb{\Gamma}$ at $t=0$, we can calculate $\mathbb{\Gamma}(t)$ for $0<t\le T$
by integrating the underlying differential equation.
The calculated $\mathbb{\Gamma}(t)$ can be used to evaluate Eq.~\eqref{eq:Sigma_def}.
Therefore, the dissipation specified by the initial point ($\mathbb{\Gamma}$ at $t=0$) is represented by
$\Sigma_0(\mathbb{\Gamma})$.
In contrast, given a point $\mathbb{\Gamma}$ at $t=T$, we can integrate the differential equation
 in reverse time to obtain $\mathbb{\Gamma}(t)$ for $0\le t< T$,
which can be used to calculate the dissipation.
The dissipation specified by the end point ($\mathbb{\Gamma}$ at $t=T$) is represented by
$\Sigma_T(\mathbb{\Gamma})$.

\begin{figure}
\begin{centering}
\includegraphics[width=13cm]{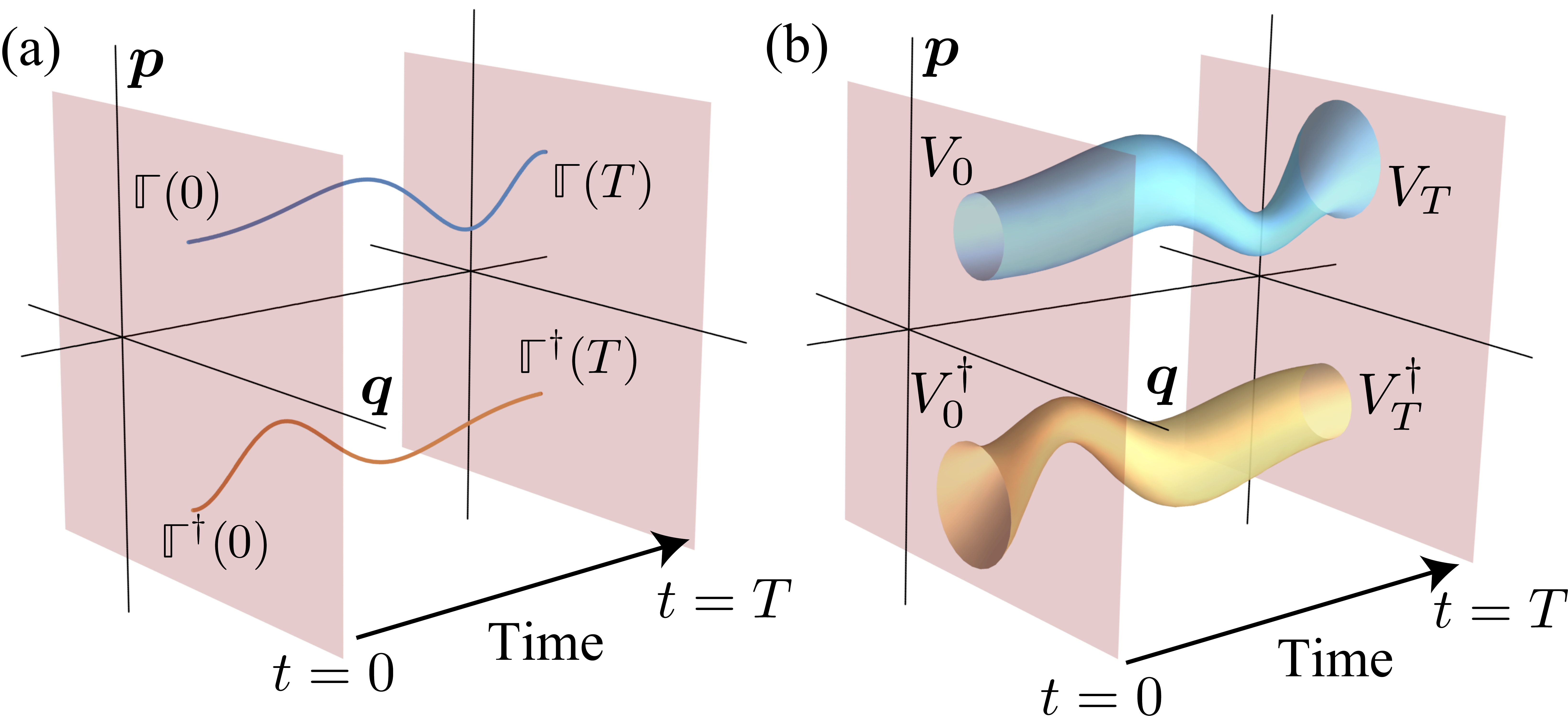}
\par\end{centering}
\caption{(a) Time evolution of $\mathbb{\Gamma}(t)$ and $\mathbb{\Gamma}^{\dagger}(t)$.
(b) Phase volume $V_0$ and $V_T$ and their time-reversal. Here,
$V_T$ is the time evolution of $V_0$, and $V_0^\dagger$ is
the time-reversal of $V_T$. Moreover, $V_T^\dagger$ is  the time evolution
of $V_0^\dagger$.
In (a) and (b), we consider a particular case $\mathbb{\Gamma}^\dagger(t) = [\bm{q}(T-t), -\bm{p}(T-t)]$ for
$\mathbb{\Gamma}(t)=[\bm{q}(t), \bm{p}(t)]$.
\label{fig:phase_space}}
\end{figure}

We next introduce a phase volume $V_{0}$ in the phase space,
which is defined such that
\begin{align}
\Sigma_0(\mathbb{\Gamma}) & =a\hspace*{1em}(\text{for any}\,\mathbb{\Gamma}\in V_{0}),\label{eq:V0_def1}\\
\Sigma_0(\mathbb{\Gamma}) & \ne a\hspace*{1em}(\text{for any}\,\mathbb{\Gamma}\notin V_{0}),\label{eq:V0_def2}
\end{align}
where $a$ is a real value. Therefore, the dynamics starting from
$\mathbb{\Gamma} \in V_{0}$ at time $t=0$ yields the dissipation
$\Sigma=a$ at time $t=T$.
Note that $V_0$ may be composed of multiple disconnected regions.
From the definition of $V_{0}$ {[}Eqs.~(\ref{eq:V0_def1})
and (\ref{eq:V0_def2}){]}, the probability to observe the dissipation
$\Sigma=a$ is given by
\[
P(\Sigma=a)=\int_{V_{0}}\rho(\mathbb{\Gamma},0)d\mathbb{\Gamma}.
\]
We also define a phase volume $V_{T}$, which is the
time evolution from $t=0$ to $t=T$ of the phase volume $V_{0}$ at time $t=0$ (Fig.~\ref{fig:phase_space}(b)).
Thus, $\mathbb{\Gamma}(0)\in V_{0}$ if and only if $\mathbb{\Gamma}(T)\in V_{T}$.

Since the dynamics is assumed to be reversible,
we can introduce a conjugate dynamics $\mathbb{\Gamma}^\dagger(t)$ of $\mathbb{\Gamma}(t)$ ($0\le t\le T$) defined by
\begin{align}
    \mathbb{\Gamma}^\dagger_{0:T}\equiv \mathcal{M}(\mathbb{\Gamma}_{0:T}),
\end{align}
where $\mathbb{\Gamma}_{0:T}$ is a trajectory from $t=0$ to $t=T$, and
$\mathcal{M}$ is a time-reversal mapping (for details of the mapping $\mathcal{M}$,
please see Ref.~\cite{Evans:2008:NoneqLiquid}).
In particular, $\mathcal{M}(\mathbb{\Gamma}(T)) = \mathbb{\Gamma}^\dagger(0)$ and
$\mathcal{M}(\mathbb{\Gamma}(0)) = \mathbb{\Gamma}^\dagger(T)$.
$\mathbb{\Gamma}^\dagger(t)$ is the (positive) time evolution starting from $\mathbb{\Gamma}^\dagger(0)$ at $t=0$,
which
corresponds to the negative time evolution starting from $\mathbb{\Gamma}(T)$ at $t=T$ (Fig.~\ref{fig:phase_space}(a)).
We hereafter assume the following relation for the distribution function
at time $t=0$:
\begin{equation}
\rho(\mathcal{M}(\mathbb{\Gamma}),0)=\rho(\mathbb{\Gamma},0).\label{eq:rho_assumption}
\end{equation}
The assumption of Eq.~(\ref{eq:rho_assumption}) is typically satisfied
by equilibrium distribution (but equilibrium distribution is not a
necessary condition) \cite{Evans:2008:NoneqLiquid}. By using the
assumption of Eq.~(\ref{eq:rho_assumption}),
$\Sigma$ is anti-symmetric under the time reversal:
\begin{align}
\Sigma_0(\mathbb{\Gamma}^{\dagger}(0)) & =\ln\left(\frac{\rho(\mathbb{\Gamma}^{\dagger}(0),0)}{\rho(\mathbb{\Gamma}^{\dagger}(T),0)}\right)-\int_{0}^{T}\Upsilon(\mathbb{\Gamma}^{\dagger}(s))ds=\ln\left(\frac{\rho(\mathbb{\Gamma}(T),0)}{\rho(\mathbb{\Gamma}(0),0)}\right)+\int_{0}^{T}\Upsilon(\mathbb{\Gamma}(s))ds=-\Sigma_0(\mathbb{\Gamma}(0)).\label{eq:Sigma_anti_sym}
\end{align}
When the time-reversal mapping is applied,
the integration of the phase space compression factor
has equal magnitude with opposite sign by construction of the time-reversal mapping.
We next introduce a phase volume $V_{0}^{\dagger}$:
\begin{equation}
V_{0}^{\dagger}\equiv\{\mathcal{M}(\mathbb{\Gamma})|\mathbb{\Gamma}\in V_{T}\}.\label{eq:V0dagger_def}
\end{equation}
Similarly, we also define a phase volume $V_{T}^{\dagger}$, which is the time evolution
from $t=0$ to $t=T$
of the phase volume $V_{0}^{\dagger}$
at time $t=0$ (Fig.~\ref{fig:phase_space}(b)).
From Eq.~\eqref{eq:Sigma_anti_sym},
the probability
to observe the dissipation $\Sigma=-a$ is given by
\begin{equation}
P(\Sigma=-a)=\int_{V_{0}^{\dagger}}\rho(\mathbb{\Gamma},0)d\mathbb{\Gamma}.\label{eq:P_minus_A}
\end{equation}
Since elements in the comoving volume do not emerge or disappear in the dynamics,
the following relation holds:
\begin{equation}
\int_{V_{0}^{\dagger}}\rho(\mathbb{\Gamma},0)d\mathbb{\Gamma}=\int_{V_{T}^{\dagger}}\rho(\mathbb{\Gamma},T)d\mathbb{\Gamma}.\label{eq:P_V0dag_VTdag}
\end{equation}
Using Eqs.~\eqref{eq:rho_formal_sol} and \eqref{eq:Sigma_def}, we have
\begin{equation}
\frac{\rho(\mathbb{\Gamma},T)}{\rho(\mathbb{\Gamma},0)}=e^{\Sigma_T(\mathbb{\Gamma})}.\label{eq:rho_ratio}
\end{equation}
Combining Eqs.~(\ref{eq:P_minus_A}) through (\ref{eq:rho_ratio}), we
obtain
\begin{align*}
P(\Sigma=-a) & =\int_{V_{T}^{\dagger}}\frac{\rho(\mathbb{\Gamma},T)}{\rho(\mathbb{\Gamma},0)}\rho(\mathbb{\Gamma},0)d\mathbb{\Gamma}=\int_{V_{T}^{\dagger}}e^{\Sigma_T(\mathbb{\Gamma})}\rho(\mathbb{\Gamma},0)d\mathbb{\Gamma}.
\end{align*}
Note that $\Sigma_T(\mathbb{\Gamma})=-a$ for $\mathbb{\Gamma}\in V_{T}^{\dagger}$,
because $\Sigma_0(\mathbb{\Gamma})=-a$ for $\mathbb{\Gamma}\in V_{0}^{\dagger}$ by construction.
Therefore,
\begin{align}
P(\Sigma=-a) & =e^{-a}\int_{V_{T}^{\dagger}}\rho(\mathbb{\Gamma},0)d\mathbb{\Gamma}=e^{-a}\int_{V_{0}}\rho(\mathbb{\Gamma},0)d\mathbb{\Gamma}=e^{-a}P(\Sigma=a),\label{eq:FT_result}
\end{align}
where we used Eq.~(\ref{eq:rho_assumption}) in the second equality.
Equation~(\ref{eq:FT_result}) is the fluctuation theorem for
the deterministic dynamical ensembles.
The derivation requires that the dynamics is reversible
so that the conjugate dynamics exists, as detailed
in Ref.~\cite{Evans:2008:NoneqLiquid}.
Moreover, the initial ensemble and
the subsequent dynamics should be \emph{ergodically consistent}. This means that,
for all possible initial phase points, the time reversal of their end points
should be included in the initial ensemble. For instance, when the initial ensemble
is isoenergetic (microcanonical) and the subsequent dynamics is
\emph{not} isoenergetic, time reversal of the endpoint may not be included
in the initial ensemble. This case is not allowed in the fluctuation
theorem.
The above derivation is a review of the proof in Ref.~\cite{Searles:2000:GeneralTFT}.

A derivation of the fluctuation theorem for a two variable case is straightforward.
We consider another observable $\Phi$. Similar to $\Sigma$,
we can specify $\Phi$ either by the initial or end points,
which we denote by $\Phi_0(\mathbb{\Gamma})$ or $\Phi_T(\mathbb{\Gamma})$, respectively.
We assume that $\Phi$ is anti-symmetric under time reversal:
\begin{equation}
\Phi_0(\mathbb{\Gamma}(0))=-\Phi_0(\mathbb{\Gamma}^{\dagger}(0)),\label{eq:Phi_antisym}
\end{equation}
which is similar to the generalized dissipation $\Sigma$.
Again we
define a phase volume $W_{0}$ in the phase space analogous to $V_{0}$:
\begin{align}
\Sigma_0(\mathbb{\Gamma}) & =a\wedge\Phi_0(\mathbb{\Gamma})=b\hspace*{1em}(\text{for any}\,\mathbb{\Gamma}\in W_{0}),\label{eq:W0_def1}\\
\Sigma_0(\mathbb{\Gamma}) & \ne a\vee\Phi_0(\mathbb{\Gamma})\ne b\hspace*{1em}(\text{for any}\,\mathbb{\Gamma}\notin W_{0}).\label{eq:W0_def2}
\end{align}
Similar to $V_0$, let $W_T$ be the time evolution of $W_0$, let $W_0^\dagger$ be the time reversal of
$W_T$ [cf. Eq.~\eqref{eq:V0dagger_def}], and let $W_T^\dagger$ be the time evolution of $W_0^\dagger$
(they are related as shown in Fig.~\ref{fig:phase_space}(b), where $V$ is replaced by $W$).
From Eqs.~\eqref{eq:W0_def1} and \eqref{eq:W0_def2},
the probability to observe $\Sigma=a$ and $\Phi = b$ is given by
\begin{align*}
P\left(\Sigma=a,\Phi=b\right) & =\int_{W_{0}}\rho(\mathbb{\Gamma},0)d\mathbb{\Gamma}.
\end{align*}
From Eqs.~\eqref{eq:Sigma_anti_sym} and \eqref{eq:Phi_antisym},
the probability to observe $\Sigma=-a$ and $\Phi = -b$ is (cf. Eq.~\eqref{eq:P_minus_A})
\begin{align*}
P\left(\Sigma=-a,\Phi=-b\right) & =\int_{W_{0}^{\dagger}}\rho(\mathbb{\Gamma},0)d\mathbb{\Gamma}.
\end{align*}
We can follow the same procedure as the previous fluctuation theorem.
We only need to replace $V_0$, $V_T$, $V_0^\dagger$, and $V_T^\dagger$ with
$W_0$, $W_T$, $W_0^\dagger$, and $W_T^\dagger$, respectively, in the derivation. Then we
obtain the fluctuation theorem with respect to $\Sigma$ and $\Phi$:
\begin{equation}
\frac{P(\Sigma=a,\Phi=b)}{P(\Sigma=-a,\Phi=-b)}=e^{a},\label{eq:deterministic_joint_FT}
\end{equation}
which is the fluctuation theorem of Eq.~\FTUsigmaUwUdef{} for the deterministic dynamical ensembles.

\subsection{Numerical verification}

We test the fluctuation theorem of Eq.~\eqref{eq:deterministic_joint_FT}
by the SLLOD equations for the planar Couette flow \cite{Searles:2000:GeneralTFT, Evans:2008:NoneqLiquid}.
This model is a particular case of the dynamical ensemble discussed above.
The system consists of $N$ particles in two dimensions
which undergo isokinetic (isothermal) shear flow.
Let $q_{xi}$ and $q_{yi}$ be the $x$ and $y$ coordinates of the $i$th particle, respectively, and
$p_{xi}$ and $p_{yi}$ be $x$ and $y$ momenta of the $i$th particle, respectively.
We define $\bm{q}_i \equiv [q_{xi},q_{yi}]$ and $\bm{q}\equiv [\bm{q}_1,\bm{q}_2,..,\bm{q}_N]$
($\bm{p}_i$ and $\bm{p}$ are defined analogously).
The equations of motion are given by the following ordinary differential equations:
\begin{align}
    \dot{\bm{q}}_i & = \frac{\bm{p}_i}{m} + \bm{u}_x \gamma q_{yi},\label{eq:sllod_qi}\\
    \dot{\bm{p}}_i & = \bm{F}_i(\bm{q}) - \bm{u}_x \gamma p_{yi} - \alpha_K(\bm{q},\bm{p})\bm{p}_i,\label{eq:sllod_pi}
\end{align}
where $m$ is the particle mass, $\bm{F}_i$ is the applied force due to the potential energy of $N$-interacting particles, $\gamma$ is the strain rate, $\bm{u}_x$ is the unit vector in $x$-direction, and $\alpha_K$ is the Gaussian
isokinetic thermostat multiplier. $\alpha_K$ is defined by
\begin{equation}
    \alpha_K(\bm{q},\bm{p}) \equiv \frac{\sum_{i=1}^N \left [ \bm{F}_i(\bm{q})\cdot\bm{p}_i - \gamma p_{xi}p_{yi} \right ] }
    {\sum_{i=1}^N \bm{p}_i \cdot \bm{p}_i},\label{eq:alpha_K_def}
\end{equation}
where $\cdot$ denotes the inner product.
We employ the Lees--Edwards boundary condition.
The internal energy is given by
\begin{equation}
    H_0 \equiv \sum_{i=1}^N \frac{\bm{p}_i \cdot \bm{p}_i}{2m} + \mathcal{U}(\bm{q}),\label{eq:H0_sllod_def}
\end{equation}
where $\mathcal{U}(\bm{q})$ is the potential energy of $N$-interacting particles
(we employ the Weeks--Chandler--Andersen potential).
The dissipation of this system is \cite{Searles:2000:GeneralTFT}
\begin{equation}
    \Sigma \simeq \frac{1}{\mathcal{T}}\left[ \mathcal{U}(\bm{q}(T)) - \mathcal{U}(\bm{q}(0)) \right ]
    + 2N \int_0^T dt\;  \alpha_K(\bm{q}(t),\bm{p}(t)),
    \label{eq:Sigma_sllod_def}
\end{equation}
where $\mathcal{T} = ({1}/{N})\sum_{i=1}^{N} {\bm{p}_i\cdot \bm{p}_i}/{(2m)}$ is
the temperature.
The time-reversal mapping of this system is \cite{Searles:2000:GKfromFT, Evans:2008:NoneqLiquid}
\begin{equation}
    \mathcal{M}[\bm{q}_{x},\bm{q}_{y},\bm{p}_{x},\bm{p}_{y}] = [\bm{q}_{x},-\bm{q}_{y},-\bm{p}_{x},\bm{p}_{y}],\label{eq:time_reversal_sllod}
\end{equation}
where $\bm{q}_x = [q_{x1},q_{x2},...,q_{xN}]$ ($\bm{q}_y$, $\bm{p}_x$, and $\bm{p}_y$ are defined analogously).
$\mathcal{M}$ of Eq.~\eqref{eq:time_reversal_sllod} is known as the Kawasaki mapping.
For an observable, we consider the kinetic energy difference in the $x$-direction:
\begin{equation}
    \Phi =  \sum_{i=1}^{N}\frac{p_{xi}(T)^{2}}{2m}-\sum_{i=1}^{N}\frac{p_{xi}(0)^{2}}{2m},\label{eq:Phi_sllod_def}
\end{equation}
which negates the sign under time reversal.

For a numerical simulation, we employ the following settings.
The number of particles is $N=16$, the temperature is $\mathcal{T} =0.5$, the strain rate is $\gamma = 0.5$, the size of the unit cell
is $8\times5$, and the time duration is $T=1.0$.
We employ the reduced units.
We solve Eqs.~\eqref{eq:sllod_qi} and \eqref{eq:sllod_pi} a total of $2\times 10^5$ times to calculate the probability density $P(\Sigma,\Phi)$.
We use the kernel density estimator to calculate the probability density function from the obtained data.

\begin{figure}
\begin{centering}
\includegraphics[width=16cm]{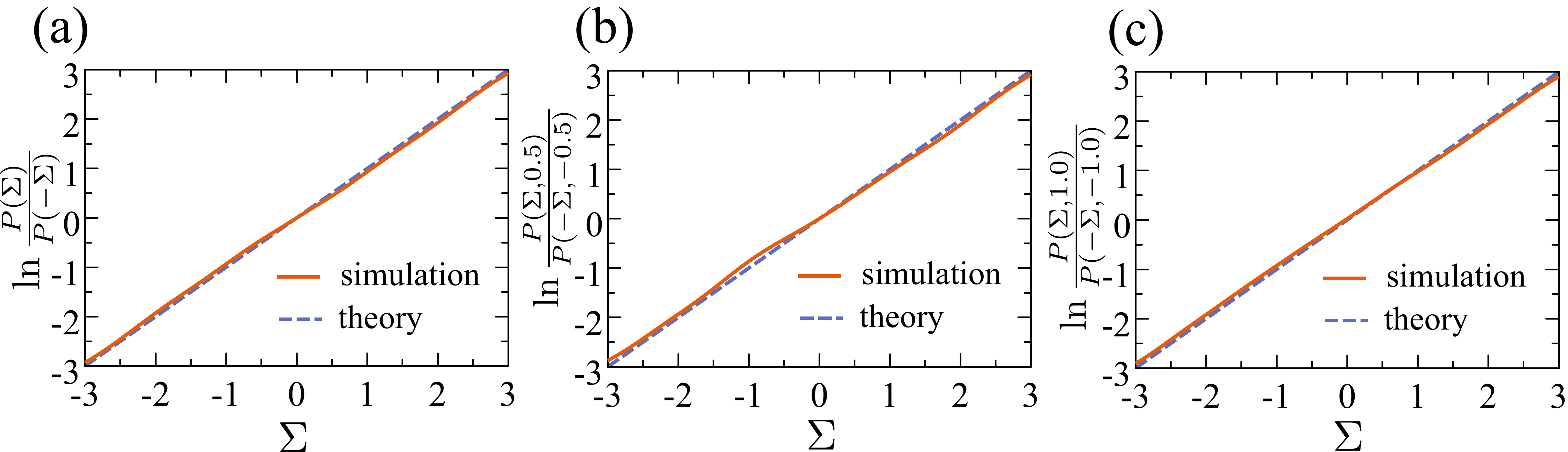}
\par\end{centering}
\caption{Verification of the fluctuation theorem for the planar Couette flow.
(a) Fluctuation theorem with respect to $\Sigma$ [Eq.~\eqref{eq:FT_result}].
The solid line denotes $\ln \left[ {P(\Sigma)}/{P(-\Sigma)}\right]$ as a
function of $\Sigma$ calculated by the computer simulation. (b) and (c) Fluctuation theorem
with respect to $\Sigma$ and $\Phi$ [Eq.~\eqref{eq:deterministic_joint_FT}].
The solid lines denote (b) $\ln \left[ {P(\Sigma,\Phi=0.5)}/{P(-\Sigma,\Phi=-0.5)} \right]$ and
(c) $\ln\left[ {P(\Sigma,\Phi=1.0)}/{P(-\Sigma,\Phi=-1.0)}\right]$ as functions of $\Sigma$
calculated by the computer simulation.
In (a)--(c), the dashed lines are theoretically expected results.
The kernel density estimator is applied to the obtained points to
calculate $P(\Sigma,\Phi)$.
\label{fig:joint_FT}}
\end{figure}

We first test the fluctuation theorem with respect to $\Sigma$ [Eq.~\eqref{eq:FT_result}], whose
satisfaction in the isokinetic planar Couette flow was already shown in Ref.~\cite{Searles:2000:GeneralTFT}.
Figure~\ref{fig:joint_FT}(a)
shows $\ln \left[ {P(\Sigma)} / {P(-\Sigma)} \right ]$ as a
function of $\Sigma$, where the solid and dashed lines are a result obtained by
the computer simulation and a theoretically expected result, respectively.
We can see a good agreement between them, reverifying the
fluctuation theorem of Eq.~\eqref{eq:FT_result}.
Next, we test the fluctuation theorem with respect to $\Sigma$ and $\Phi$ [Eq.~\eqref{eq:deterministic_joint_FT}].
We calculate $\ln \left [ {P(\Sigma,\Phi=0.5)}/{P(-\Sigma,\Phi=-0.5)} \right ]$
and $\ln \left[ {P(\Sigma,\Phi=1.0)}/{P(-\Sigma,\Phi=-1.0)} \right]$ in
Figs.~\ref{fig:joint_FT}(b) and (c), respectively. The meanings of the solid and dashed lines
are identical to those in Fig.~\ref{fig:joint_FT}(a).
For both Figs.~\ref{fig:joint_FT}(b) and (c), we confirm that
the results of computer simulations agree well with theoretically expected results,
which verifies the fluctuation theorem of Eq.~\eqref{eq:deterministic_joint_FT}
in the dynamical ensemble.

\section{Examples \label{sec:examples}}

\subsection{Example 1: Brownian particle on a ring}
In the main text, we consider the following Langevin equation:
\begin{equation}
    \frac{dx}{dt} = A(x) + \sqrt{2D}\xi(t),\label{eq:periodic_Langevin_def}
\end{equation}
where $A(x)$ is a drift term,
$D$ is the noise intensity,
and $\xi(t)$
is the white Gaussian noise with $\langle \xi(t) \rangle = 0$ and $\langle \xi(t)\xi(t') \rangle = \delta(t-t')$.
For analytical calculation, we consider the $A(x)=f$ case, where $f$ is a constant force applied to the particle.
Since Eq.~\eqref{eq:periodic_Langevin_def}
is a simple diffusion equation,
when considering the ring on the infinite line,
its time-dependent probability density $P(x,t)$ given $x(0)=0$, is expressed as
\begin{equation}
    P(x,t)=\frac{1}{\sqrt{4\pi Dt}}\exp\left(-\frac{(x-ft)^{2}}{4Dt}\right).\label{eq:Pxt_def}
\end{equation}
In the main text, we consider the following observables:
\begin{align}
    \phi_{\mathrm{sgn}}(\Gamma)&\equiv\mathrm{sign}\left(\jmath(\Gamma)\right)=\mathrm{sign}\left(x(T)-x(0)\right),\label{eq:sg_def}
\end{align}
where $\jmath(\Gamma) \equiv \int_0^T \dot{x}dt$.
For $\phi_\mathrm{sgn}$,
let $\bms{P}(\phi_\mathrm{sgn},t)$ be the probability of $\phi_\mathrm{sgn} \in \{-1,1\}$ at time $t$.
Then, $\bms{P}(\phi_\mathrm{sgn},t)$ is given by
\begin{align}
    \bms{P}(1,t)&=\int_{0}^{\infty}dx\,P(x,t)=\frac{1}{2}\left[1+\mathrm{erf}\left(\frac{f}{2}\sqrt{\frac{t}{D}}\right)\right]\label{eq:Psg1_def}\\
    \bms{P}(-1,t)&=\int_{-\infty}^{0}dx\,P(x,t)=\frac{1}{2}\left[1-\mathrm{erf}\left(\frac{f}{2}\sqrt{\frac{t}{D}}\right)\right].\label{eq:Psg2_def}
\end{align}
Using Eqs.~\eqref{eq:Psg1_def} and \eqref{eq:Psg2_def}, the mean and the variance of $\phi_\mathrm{sgn}$ are
\begin{align}
    \left\langle \phi_{\mathrm{sgn}}\right\rangle &=\sum_{\phi_{\mathrm{sgn}}=\pm1}\phi_{\mathrm{sgn}}\bms{P}(\phi_{\mathrm{sgn}},t)=\mathrm{erf}\left(\frac{f}{2}\sqrt{\frac{t}{D}}\right),\label{eq:mean_phi_sg}\\
    \mathrm{Var}[\phi_{\mathrm{sgn}}]&=\sum_{\phi_{\mathrm{sgn}}=\pm1}\phi_{\mathrm{sgn}}^{2}\bms{P}(\phi_{\mathrm{sgn}},t)
    -\langle \phi_\mathrm{sgn} \rangle^2
    =1-\mathrm{erf}\left(\frac{f}{2}\sqrt{\frac{t}{D}}\right)^{2}.\label{eq:var_phi_sg}
\end{align}
In the steady state, the entropy production from $t=0$ to $t=T$ is given by \cite{Seifert:2012:FTReview}
\begin{equation}
\left\langle \sigma\right\rangle =T\int_0^1 dx\,\frac{A(x)J^{\mathrm{ss}}}{D}=\frac{Tf^{2}}{D},\label{eq:entprod_def}
\end{equation}
where $J^\mathrm{ss}$ is the steady-state probability current.
Using Eqs.~\eqref{eq:mean_phi_sg} through \eqref{eq:entprod_def}, we obtain
\begin{align}
    \frac{\mathrm{Var}[\phi_{\mathrm{sgn}}]}{\left\langle \phi_{\mathrm{sgn}}\right\rangle ^{2}}&=-1+\mathrm{erf}\left(\frac{\sqrt{\left\langle \sigma\right\rangle }}{2}\right)^{-2},\label{eq:phisg_ineq}
\end{align}
which is given as Eq.~\ratioUsign{} in the main text.

\begin{figure}
\centering{}\includegraphics[width=13cm]{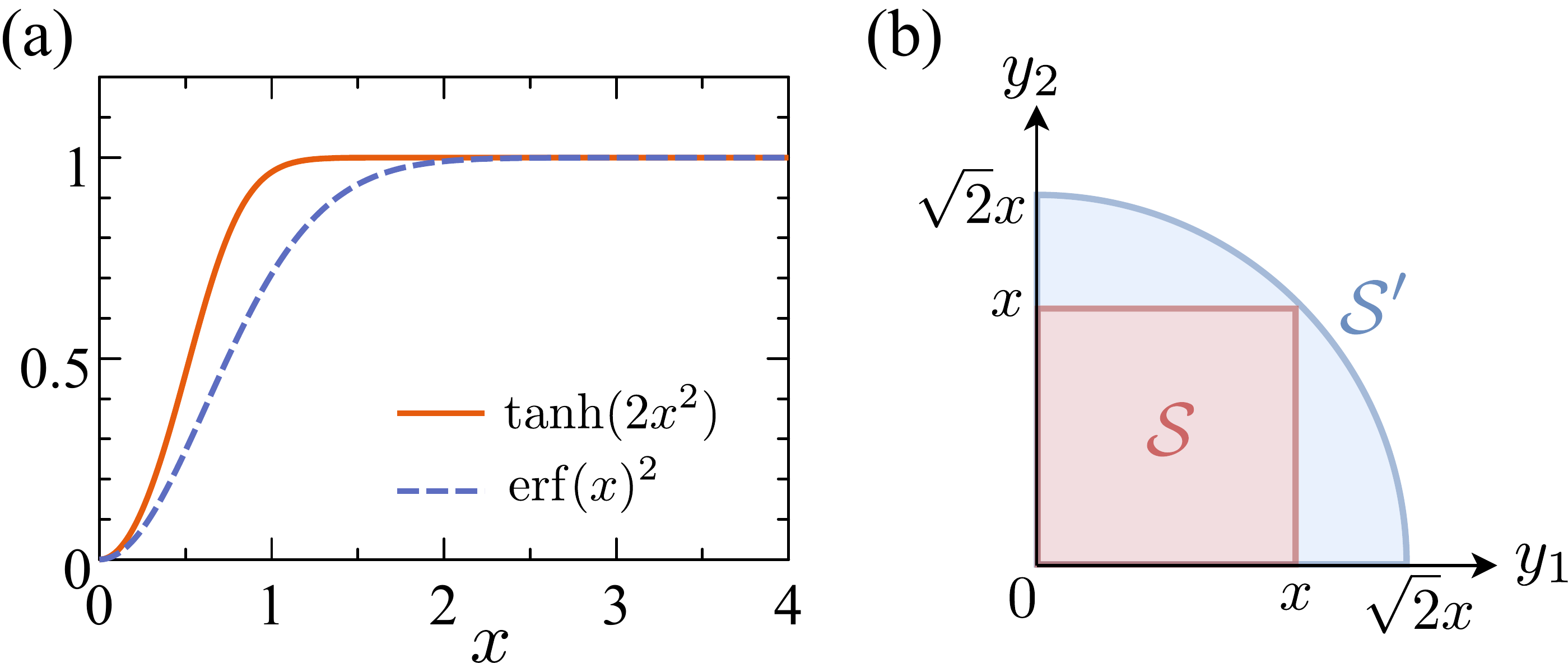} \caption{(a) Comparison of ${\tanh(2x^{2})}$ and $\mathrm{erf}(x)^2$,
which are depicted by solid and dashed lines, respectively.
(b) Relation between $\mathcal{S}$ and $\mathcal{S}^\prime$ considered in
the integration of Eq.~\eqref{eq:tanh_erf_ineq}.
\label{fig:tanh_erf}}
\end{figure}

In the main text, we mentioned that the right-hand side of Eq.~\eqref{eq:phisg_ineq}
is larger than the lower bound of Eq.~\mainUresult{}, i.e.,
\begin{equation}
-1+\mathrm{erf}\left(\frac{\sqrt{\left\langle \sigma\right\rangle }}{2}\right)^{-2}\ge\frac{2}{e^{\left\langle \sigma\right\rangle }-1}.\label{eq:erf_tanh_def}
\end{equation}
In order to show the relation of Eq.~\eqref{eq:erf_tanh_def}, it
is sufficient to show the following relation:
\begin{equation}
\tanh(2x^{2})\ge\mathrm{erf}(x)^{2}\hspace*{1em}(x\ge0),\label{eq:main_ineq}
\end{equation}
which is obvious when evaluating both sides numerically (Fig.~\ref{fig:tanh_erf}(a)).
We can prove Eq.~\eqref{eq:main_ineq} as follows:
\begin{align}
\mathrm{erf}(x)^{2} & =\left(\frac{2}{\sqrt{\pi}}\int_{0}^{x}dy\,e^{-y^{2}}\right)^{2}\nonumber \\
 & =\frac{4}{\pi}\iint_{\mathcal{S}}dy_{1}dy_{2}\,e^{-y_{1}^{2}-y_{2}^{2}}\hspace*{1em}\left(\mathcal{S}\equiv\left\{ [y_{1},y_{2}]\Bigr|0\le y_{1}\le x,0\le y_{2}\le x\right\} \right)\nonumber \\
 & \le\frac{4}{\pi}\iint_{\mathcal{S}^{\prime}}dy_{1}dy_{2}\,e^{-y_{1}^{2}-y_{2}^{2}}\hspace*{1em}\left(\mathcal{S}^{\prime}\equiv\left\{ [y_{1},y_{2}]\Bigr|0\le\sqrt{y_{1}^{2}+y_{2}^{2}}\le\sqrt{2}x,0\le y_{1},0\le y_{2}\right\} \right)\nonumber \\
 & =\frac{4}{\pi}\int_{0}^{\frac{\pi}{2}}d\theta\int_{0}^{\sqrt{2}x}dr\,re^{-r^{2}}\nonumber \\
 & =1-e^{-2x^{2}}\nonumber \\
 & \le\tanh(2x^{2}).\label{eq:tanh_erf_ineq}
\end{align}
The third line follows because $\mathcal{S} \subseteq \mathcal{S}^{\prime}$
(we show $\mathcal{S}$ and $\mathcal{S}^{\prime}$ in Fig.~\ref{fig:tanh_erf}(b)).
The final inequality can be shown by
\[
\tanh(2x^{2})-(1-e^{-2x^{2}})=\frac{(1-e^{-2x^{2}})^{2}}{e^{2x^{2}}+e^{-2x^{2}}}\ge0.
\]

\subsection{Example 2: Dragged Brownian particle}

In the main text, we consider the following Langevin equation:
\begin{align}
\frac{dx}{dt} & =-\frac{\partial}{\partial x}U(x,\lambda)+\sqrt{2D}\xi(t),\label{eq:dragged_Langevin_def}
\end{align}
where $U(x,\lambda)$ is a potential function $U(x,\lambda)\equiv\beta(x-\lambda)^{2}/2$
with a model parameter $\beta>0$, and $\xi(t)$ and $D$ are defined
in Eq.~\eqref{eq:periodic_Langevin_def}. Here, $\lambda(t)$ is
a protocol defined as
\begin{equation}
\lambda(t)=\begin{cases}
{\displaystyle \frac{2\ell}{T}t} & {\displaystyle 0\le t<\frac{T}{2}}\\
{\displaystyle -\frac{2\ell}{T}t+2\ell} & {\displaystyle \frac{T}{2}\le t\le T}
\end{cases},\label{eq:protocol_def}
\end{equation}
where $\ell$ is the height of the signal (Fig.~\FIGmodel{}(c)).
Since Eq.~\eqref{eq:dragged_Langevin_def} reduces to the Ornstein--Uhlenbeck
process, by using the moment method (see Ref.~\cite{Tuckwell:2009:HHeq}
for the detailed derivation), the mean $\mu(t)\equiv\left\langle x(t)\right\rangle $
and the variance $\zeta(t)\equiv\mathrm{Var}[x(t)]$ of Eq.~\eqref{eq:dragged_Langevin_def}
are expressed as
\begin{align}
\dot{\mu}(t) & =-\beta(\mu(t)-\lambda(t)),\label{eq:mu_def}\\
\dot{\zeta}(t) & =-2\beta\gamma(t)+2D.\label{eq:gamma_def}
\end{align}
Let $P(x_{T},T|x_{0},0)$ be the probability density function of $x_{T}$
at time $t=T$ starting from $x_{0}$ at $t=0$. Solving Eqs.~\eqref{eq:mu_def}
and \eqref{eq:gamma_def} with initial values $\mu(0)=x_{0}$ and
$\gamma(0)=0$, we obtain
\begin{equation}
P(x_{T},T|x_{0},0)=\mathcal{N}\left(x_{T};\frac{1}{\beta T}\left(-4e^{-\beta T/2}\ell+\left(\beta Tx_{0}+2\ell\right)e^{-T\beta}+2\ell\right),\frac{D}{\beta}\left(1-e^{-2\beta T}\right)\right),\label{eq:PxT_x0_def}
\end{equation}
where $\mathcal{N}(x;\mu,\sigma^{2})$ is a Gaussian distribution
with mean $\mu$ and variance $\sigma^{2}$. Since the system
is in equilibrium at $t=0$, we have
\begin{equation}
P(x_{0},0)=\mathcal{N}\left(x_{0};0,\frac{D}{\beta}\right).\label{eq:Px0_def}
\end{equation}
The observable considered in the main text is
\[
\phi_{\mathrm{pos}}(\Gamma)\equiv\int_{0}^{T}\dot{x}(t)dt=x(T)-x(0).
\]
Therefore, from Eqs.~\eqref{eq:PxT_x0_def} and \eqref{eq:Px0_def},
we have
\begin{align*}
\left\langle \phi_{\mathrm{pos}}\right\rangle  & =\int dx_{T}\int dx_{0}P(x_{T},T|x_{0},0)P(x_{0},0)(x_{T}-x_{0}),\\
 & =\frac{2\ell\left(e^{-\beta T}-2e^{-\beta T/2}+1\right)}{\beta T},\\
\mathrm{Var}[\phi_{\mathrm{pos}}] & =\int dx_{T}\int dx_{0}P(x_{T},T|x_{0},0)P(x_{0},0)(x_{T}-x_{0})^{2}-\left\langle \phi_{\mathrm{pos}}\right\rangle ^{2},\\
 & =\frac{2D\left(1-e^{-\beta T}\right)}{\beta}.
\end{align*}
and thus
\begin{equation}
\frac{\mathrm{Var}[\phi_{\mathrm{pos}}]}{\left\langle \phi_{\mathrm{pos}}\right\rangle ^{2}}=\frac{D\beta T^{2}\left(1-e^{-\beta T}\right)}{2\ell^{2}\left(e^{-\beta T}-2e^{-\beta T/2}+1\right)^{2}},\label{eq:ratio_def}
\end{equation}
which is Eq.~\varUmeanUdragged{} in the main text. The work exerted
on the particle is \cite{Seifert:2012:FTReview}
\begin{align}
w(\Gamma) & =\int_{0}^{T}dt\,\frac{\partial U(x,\lambda)}{\partial\lambda}\frac{d\lambda}{dt}\nonumber \\
 & =-\int_{0}^{T}dt\,\beta(x(t)-\lambda(t))\frac{d\lambda}{dt}.\label{eq:work_def}
\end{align}
Taking the average of Eq.~\eqref{eq:work_def}, we obtain
\begin{align}
\left\langle w\right\rangle  & =-\int_{0}^{T}dt\,\beta(\mu(t)-\lambda(t))\frac{d\lambda}{dt}\nonumber \\
 & =\frac{4\ell^{2}\left(-e^{-\beta T}+\beta T+4e^{-\beta T/2}-3\right)}{\beta T^{2}},\label{eq:w_ave_def}
\end{align}
where $\mu(t)$ is a solution of Eq.~\eqref{eq:mu_def} with an initial
value of $\mu(0)=0$. Equation~\eqref{eq:w_ave_def} is denoted as Eq.~\wUaverage{}
in the main text.

Using Eqs.~\eqref{eq:ratio_def} and \eqref{eq:w_ave_def}, we next
prove the following inequality considered in the main text:
\begin{equation}
\frac{\mathrm{Var}[\phi_{\mathrm{pos}}]}{\left\langle \phi_{\mathrm{pos}}\right\rangle ^{2}}\ge\frac{2}{\left\langle w\right\rangle /D}.\label{eq:example2_ineq_def}
\end{equation}
From Eq.~\eqref{eq:w_ave_def}, $\left\langle w\right\rangle >0$
for $\beta>0$ and $T>0$. Therefore, defining a variable $\kappa\equiv\beta T$,
the inequality of Eq.~\eqref{eq:example2_ineq_def} is calculated
as
\begin{equation}
\frac{\left(1-e^{-\kappa}\right)\left(-e^{-\kappa}+\kappa+4e^{-\kappa/2}-3\right)}{\left(e^{-\kappa}-2e^{-\kappa/2}+1\right)^{2}}\ge1\hspace*{1em}(\text{for }\kappa>0).\label{eq:example2_ineq2}
\end{equation}
From Eq.~\eqref{eq:example2_ineq2}, we need to prove $\mathcal{K}(\kappa)\ge0$ for $\kappa \ge 0$,
where
\begin{equation}
\mathcal{K}(\kappa)\equiv\left(1-e^{-\kappa}\right)\left(-e^{-\kappa}+\kappa+4e^{-\kappa/2}-3\right)-\left(e^{-\kappa}-2e^{-\kappa/2}+1\right)^{2}.\label{eq:Uk_def}
\end{equation}
Since $\mathcal{K}'(\kappa)=e^{-\kappa}(\kappa-4e^{\kappa/2}+e^{\kappa}+3)\ge0$
(note that $\partial_{\kappa}\left(\kappa-4e^{\kappa/2}+e^{\kappa}+3\right)=(e^{\kappa/2}-1)^{2}\ge0$),
we obtain $\mathcal{K}^{\prime}(\kappa)\ge0$. Using $\mathcal{K}(0)=0$,
we can conclude that $\mathcal{K}(\kappa)\ge0$ for $\kappa\ge0$,
which proves Eq.~\eqref{eq:example2_ineq_def}. Moreover, the inequality
of Eq.~\eqref{eq:example2_ineq_def} saturates when $\beta T\to0$.

\subsection{Example 3: Discrete-time random walk on a ring}

\begin{figure}
\centering{}\includegraphics[width=11cm]{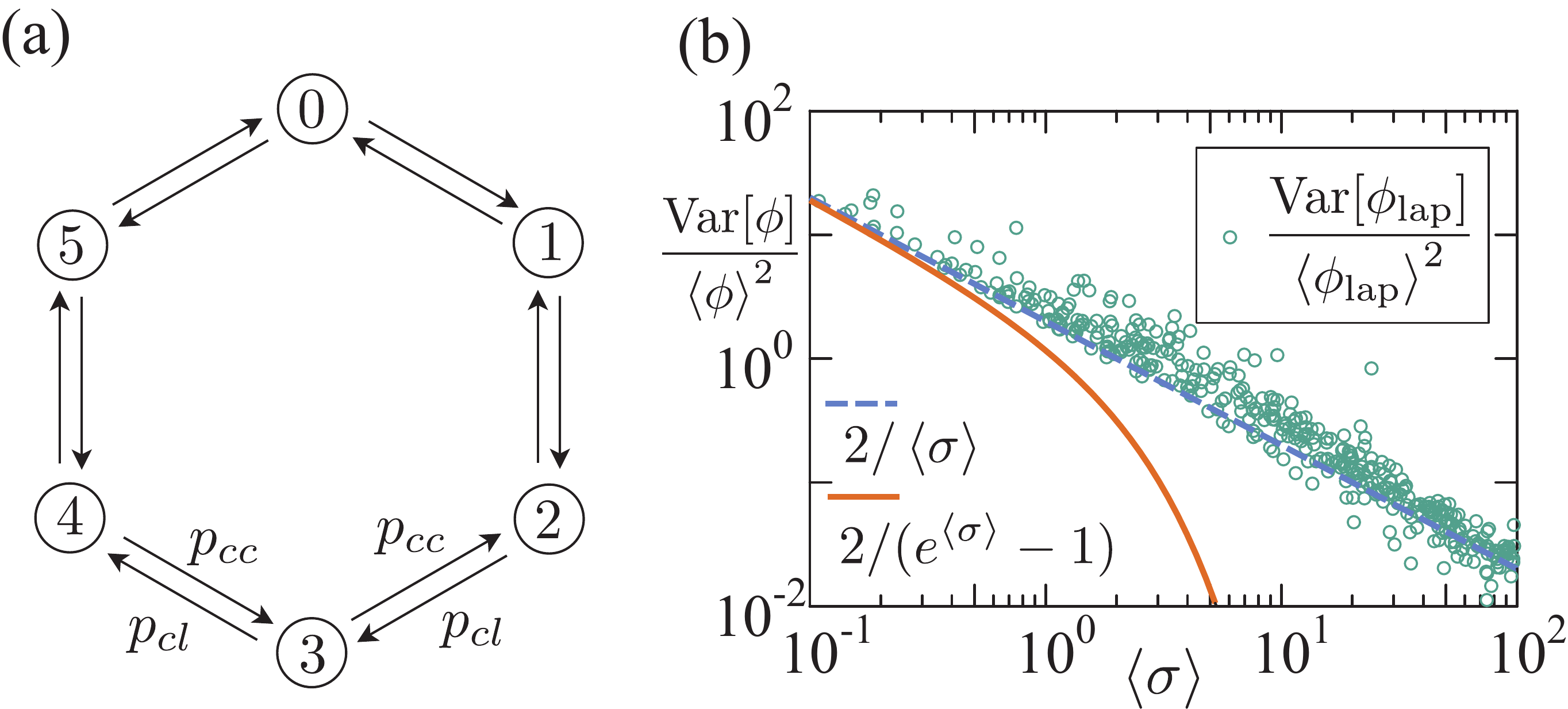}
\caption{(a) Random walk on the ring for the $L=6$ case. The transition probabilities of the clockwise and
counter clockwise directions are $p_{cl}$ and $p_{cc}$, respectively. (b)
$\mathrm{Var}[\phi_\mathrm{lap}]/\langle \phi_\mathrm{lap} \rangle^2$ and the lower bounds of the FTUR and TUR as a
function of $\langle \sigma \rangle$. The lower bounds of Eq.~\mainUresult{} and Eq.~\TURUcontinuous{}
are depicted
by solid and dashed lines, respectively. Circles denote $\mathrm{Var}[\phi_\mathrm{lap}]/\langle \phi_\mathrm{lap} \rangle^2$ computed by the computer simulation, where $p_{cl}$, $p_{cc}$, $L$, and $T$ are randomly selected
as $p_{cl}\in (0,1)$, $p_{cc}\in (0,1)$, $L\in \{1,2,\cdots,20\}$, $T=k_L L$ with $k_L = \{5,6,\cdots,20\}$.
\label{fig:randomwalk_laps}}
\end{figure}

We consider the uncertainty relation for the number of laps in
a discrete-time random walk.
The discrete-time random walk is defined on a ring with $L$ states (Fig.~\ref{fig:randomwalk_laps}(a)).
Let $p_{cc}$ and $p_{cl}$ be the probabilities of jumping
to neighbor states in the clockwise and counter-clockwise directions, respectively, at each step
($p_{cc}>0$, $p_{cl} > 0$, and $p_{cc}+p_{cl}\le 1$).
Let $x(t)$ be the equivalent position of the random walker at step $t$ on an infinite line, i.e., $x(t) \in \{\cdots,-2,-1,0,1,2,\cdots \}$.
We consider an observable which counts the number of laps:
\begin{equation}
    \phi_\mathrm{lap}(\Gamma) \equiv \mathrm{quotient}\left [ x(T) - x(0), L \right ],
\end{equation}
where $\Gamma$ is a trajectory, $\Gamma \equiv [x(0),x(1),\cdots x(T)]$ ($T$ is a maximum step), and
$\mathrm{quotient}(x,y)$ is the quotient function.
For $x \ge 0$ and $y > 0$, $\mathrm{quotient}(x, y)$ gives the
quotient of $x$ divided by $y$, e.g., $\mathrm{quotient}(7,2) = 3$.
For $x < 0$ and $y > 0$, $\mathrm{quotient}(x,y)$ works symmetrically to the positive case, e.g.,
$\mathrm{quotient}(-7, 2) = -3$.
Therefore, $\mathrm{quotient}(x,y)$ is an odd function of $x$.
When the random walker laps the ring in the clockwise (counter-clockwise) direction,
the lap is incremented (decremented) by $1$.
Since $\phi_\mathrm{lap}(\Gamma^\dagger) = -\phi_\mathrm{lap}(\Gamma)$,
it should satisfy the FTUR (Eq.~\mainUresult{}).

We verify the FTUR of Eq.~\mainUresult{} for the discrete-time random walk by a computer simulation.
We randomly select $L$, $T$, $p_{cc}$, and $p_{cl}$, and calculate
$\mathrm{Var}[\phi_\mathrm{lap}] / \langle \phi_\mathrm{lap} \rangle^2$ and $\langle \sigma \rangle$ for the selected parameter values as the average of $10^6$ trajectories (the range of the
parameters is shown in the caption of Fig.~\ref{fig:randomwalk_laps}(b)).
The calculated
realizations are shown by circles in Fig.~\ref{fig:randomwalk_laps}(b).
The lower bounds of Eq.~\mainUresult{}
and Eq.~\TURUcontinuous{} are shown by solid and dashed lines, respectively.
Although $\mathrm{Var}[\phi_\mathrm{lap}] / \langle \phi_\mathrm{lap} \rangle^2$ is smaller than the bound of Eq.~\TURUcontinuous{} for larger $\langle \sigma \rangle$,
it is always larger than the bound of Eq.~\mainUresult{}.
In the main text, we have experimented on the continuous-time Markov chains
with observables which are anti-symmetric under time reversal.
The result of this section demonstrates that the FTUR holds for the discrete-time Markov chain with
the observable which is an odd function of the current.

